\documentclass[aps,twocolumn,pra,floatfix,preprintnumbers]{revtex4}

\usepackage{amsmath}
\usepackage{amssymb}
\usepackage{pifont}
\usepackage{graphicx}
\usepackage{hhline,times}
\usepackage{graphicx}

\newcommand{\ket}[1]{\left\vert#1\right\rangle}
\newcommand{\bra}[1]{\left\langle#1\right\vert}
\newcommand{\beq}{\begin{equation}}
\newcommand{\eeq}{\end{equation}}
\newcommand{\bea}{\begin{eqnarray}}
\newcommand{\eea}{\end{eqnarray}}
\newcommand{\inn}[2]{\left\langle#1\vert#2\right \rangle}

\makeatletter
\def\btt#1{\texttt{\@backslashchar#1}}
\DeclareRobustCommand\bblash{\btt{\@backslashchar}}
\makeatother

\begin{document}

\title{Quantum-Secured Surveillance Based on Mach-Zehnder Interferometry} 

\author{C. Allen Bishop$^{1,}$\footnote{Email: bishopca@ornl.gov}, Travis S. Humble$^{1}$, Ryan S. Bennink$^{1}$, 
and Brian P. Williams$^{1,2}$}
\affiliation{$^{1}$ Quantum Information Science Group, Computational Sciences and Engineering Division, 
Oak Ridge National Laboratory, Oak Ridge, Tennessee 37831-6418, USA \\
$^{2}$ Department of Physics, University of Tennessee, 
Knoxville, Tennessee 37996-1200, USA}

\date{\today}

\begin{abstract}
We present a method for intrusion detection which is based on the Mach-Zehnder interference effect. This device provides monitored surveillance by continuously measuring 
the intensity of light collected by a pair of photodetectors. We find that our protocol allows for the detection of intrusion 
attempts which employ path-redirection and/or intercept-resend techniques. Expectation values for the registered output flux are provided for normal and interrupted operation.

\end{abstract}


\maketitle


\emph{Introduction.--}
Optical tripwire systems offer quick intrusion 
notification in most scenarios, e.g., fiber cutting or beam blocking intrusions. The tripwire usually consists of a laser beam, which in turn consists of a large collection of photons. This presents a vulnerability which could be exploited in a sophisticated attack. By tapping off a small portion of the original signal, an adversary can determine, with high precision, the properties inherent to 
the surveillance beam while leaving the majority of the signal unaffected \cite{Mandel,Boyd}. This can be achieved for both unmodulated and modulated waveforms. If the interrogated signal strength is sufficiently weak, the characteristics of the tripwire could be determined using direct detection and dyne detection without causing alarm. A decoy source could then be used to mimic the original signal. The decoy source could be set in place, for instance, during a forced power outage, and constructed to switch on remotely. This would provide unannounced entry and exit between the two sources at any later time.

In this paper, we present a method for intrusion detection which circumvents this issue. Our method is based on the 
Mach-Zehnder interference (MZI) effect \cite{Zehnder} and provides surveillance by continuously measuring the intensity of light collected by a pair of photodetectors. Under normal conditions, each photodetector output current per signal pulse is expected to remain constant. When an intruder passes through the secured perimeter, the output signal drops, signaling an alarm. This feature is shared with common laser tripwire designs. However, our contribution closes the intercept-resend vulnerability facing existing optical monitoring techniques by using MZI to test the integrity and authenticity of the transmitted signal. Thus, our method offers an additional layer of security relative to classical optical sensors. Like earlier efforts employing related effects \cite{Humble,Malik}, a quantum optical sensor can mitigate the redirection and intercept-resent vulnerabilities facing their classical counterparts.

\begin{figure}[!h]
\includegraphics[width=.33\textwidth]{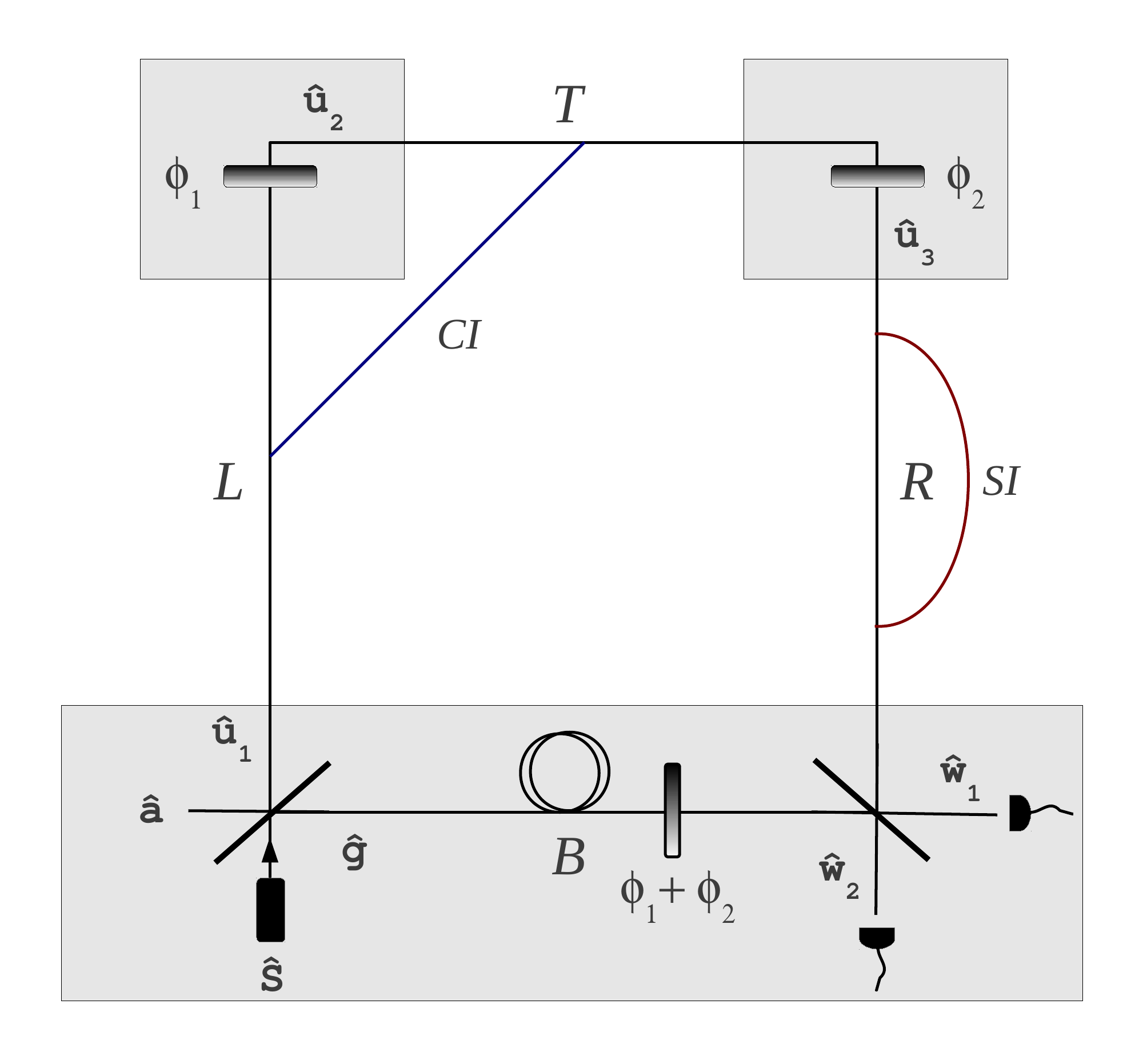}
\caption{Security protocol based on a balanced Mach-Zehnder interferometer. A {\it{side intrusion}} (SI) path-diversion attempt along the right side $R$ 
is indicated with the red line. A {\it{cross intrusion}} (CI) diversion attempt from side $L$ to $T$ is depicted with the blue line.}
\label{fig:lock}
\end{figure}

\emph{Normal operation.---}
\label{sec:noiseless}
An illustration of the sensor 
is given in Fig.~\ref{fig:lock}. The setup consists of a balanced Mach-Zehnder interferometer 
with a photon source located at port $\hat{s}$. The left $L$, top $T$, and right $R$ arms of the interferometer constitute the 
secured perimeter (i.e., the fence) while regions specified with a shaded box are assumed to be secured enclaves. Let $l_L,l_T,l_R,$ and $l_B$ denote the unperturbed 
path lengths of the lower, top, right, and bottom arms. Under normal, uninterrupted operation, $l_L+l_T+l_R=l_B.$ 

The transformation describing the action of the first beam splitter is given by 
$U_1 = \exp{\left[ i\pi({\hat{a}}^{\dagger}{\hat{s}} +{\hat{a}}{\hat{s}}^{\dagger})/4 \right]}.$ Using the Campbell-Baker-Haussdorff formula, we may solve the Heisenberg 
equations ${\hat{u}}_1=U_1^{\dagger}{\hat{s}} U_1$ and ${\hat{g}}=U_1^{\dagger}{\hat{a}} U_1$ to obtain
${\hat{u}}_1=({\hat{s}}+i{\hat{a}})/\sqrt{2}$ and ${\hat{g}}=({\hat{a}}+i{\hat{s}})/\sqrt{2}$. An input state of the form $\ket{\Psi}=\ket{{\text{vac}}}_{\hat{a}} \otimes \ket{1}_{\hat{s}}$ evolves to 
$\ket{\Psi^{\prime}}=U_1 \ket{\Psi} = (\ket{1}_{\hat{u}_1}\ket{{\text{vac}}}_{\hat{g}} -i\ket{0}_{\hat{u}_1}\ket{1}_{\hat{g}})/\sqrt{2}$. 
For the 
moment, let $\phi_1=\phi_2=0$ and ${\hat{u_1}}={\hat{u_2}}={\hat{u_3}}$. The second beam splitter is identical to the first, so the second transformation   
$U_2 = \exp{\left[i\pi({\hat{u}}_3^{\dagger}{\hat{g}} +{\hat{u}}_3{\hat{g}}^{\dagger})/4 \right]}$
yields the relations ${\hat{w}}_1=({\hat{g}}+i{\hat{u}}_3)/\sqrt{2}$ and 
${\hat{w}}_2=({\hat{u}}_3+i{\hat{g}})/\sqrt{2}$. The state which exits the interferometer is thus $U_2\ket{\Psi^{\prime}} = -i\ket{1}_{\hat{w}_1}\ket{{\text{vac}}}_{{\hat{w}}_2}$. 
If the source at $\hat{s}$ 
supplies the interferometer with single photons we should expect to observe 
\bea 
&\hdots& \ket{\Psi}_{j+3} \ket{\Psi}_{j+2} \ket{\Psi}_{j+1} \ket{\Psi}_{j} \hdots 
\nonumber \\
&\rightarrow& 
\hdots ({\hat{w}}_1)_{j+3} ({\hat{w}}_1)_{j+2} ({\hat{w}}_1)_{j+1} ({\hat{w}}_1)_{j} \hdots
\eea
In this description, $\ket{\Psi}_{1}$ represents the first photon which is emitted, 
$\ket{\Psi}_{2}$ represents the second, etc. The corresponding detection events are 
given by $({\hat{w}}_1)_{1}$, $({\hat{w}}_1)_{2}$, etc. Now suppose an intruder simply crosses the perimeter. The state of the bottom arm is mixed during the time for which the fence path is blocked. After tracing over the fence branch the state of the lower arm is described by
$\ket{\Psi}_{j} \rightarrow \rho_{j} = \frac{1}{2}(\ket{{\text{vac}}}_g \!\bra{{\text{vac}}} + \ket{1}_g \!\bra{1})$, i.e., the field in the bottom arm is in a mixed state of zero and one photons. No interference occurs at the second beam splitter during the time for which the patrol arm is blocked. We will observe a decrease in the output flux at $\hat{w}_1$ during this time. Specifically, if 
${\cal{F}}$ denotes the input flux at $\hat{s}$, the output flux at $\hat{w}_1$ during the time for which the fence is blocked is given by ${\cal{F}}/4$.

In what follows, we will assume that the sensor is supplied with photons emitted via parametric fluorescence \cite{Mandel,Ou}. Specifically, we will associate $\hat{s}$ with the signal mode of a down-conversion process. The idler 
mode $\hat{i}$ (not shown in the figure) will then serve as a heralding source for photon injection. If first-order processes dominate the interaction, a successful down-conversion can be approximated by the following two-photon state \cite{Ou} (neglecting polarization)
\beq
\ket{\Psi}_{\text{PDC}} = \int d\omega_1 d\omega_2 \Phi(\omega_1,\omega_2)\hat{s}^{\dagger}(\omega_1) \hat{i}^{\dagger}(\omega_2)\ket{\text{vac}},
\eeq
where $\Phi(\omega_1,\omega_2)$ is a suitable spectral amplitude distribution. Throughout, unless otherwise stated, the limits of integration are taken to be 
$(-\infty,\infty)$. Let $\ket{\Psi}_{\text{PDC},j}$ denote the $j$th down-converted pair. Suppose we detect the $j$th 
idler photon at time $t=t_j$. At this moment, the state of the signal field is obtained by projecting $\ket{\Psi}_{\text{PDC},j}$ onto the state \cite{Ou2}
\beq
\ket{\psi(t_j)} = {\cal{C}} \int  d\omega \hat{i}^{\dagger}(\omega) \eta(\omega) e^{2 \pi i \omega t_j} \ket{\text{vac}},
\eeq
where ${\cal{C}}$ is a normalization constant and $\eta(\omega)$ represents the real spectral transfer function of the idler detector. Conditioned on a detection of the idler at time $t=t_j$, the $j$th signal
photon is injected into the interferometer in the state 
\bea
\ket{\Psi}_{j} =  \inn{\psi(t_j)}{\Psi}_{\text{PDC},j} 
&=& {\cal{C}}^* \!\int \! d\omega_1 d\omega_2 \eta(\omega_2) e^{-2 \pi i \omega_2 t_j}\nonumber \\
&&\times \Phi(\omega_1,\omega_2)\hat{s}^{\dagger}(\omega_1) \ket{\text{vac}}. 
\eea
Here we have used the continuous mode commutation relation $\left[\hat{i}(\omega),\hat{i}^\dagger(\omega^{\prime}) \right]$ $= \delta(\omega -\omega^{\prime})$. For simplicity, we will assume a monochromatic cw pump beam of frequency $\omega_p$, i.e., $\Phi(\omega_1,\omega_2) \propto \delta(\omega_1 + \omega_2 - \omega_p)h\left[{\cal{L}}\Delta k(\omega_1,\omega_2) \right].$ The phase-matching function $h$ is given by 
\beq
h\left[{\cal{L}}\Delta k(\omega_j,\omega) \right] = e^{-i {\cal{L}}\Delta k(\omega_j,\omega)/2} \text{sinc}({\cal{L}}\Delta k(\omega_1,\omega_2)/2), 
\eeq
and ${\cal{L}}$ denotes the crystal length. For frequency degenerate processes, we can apply a first-order Taylor expansion of the phase mismatch $\Delta k(\omega_1,\omega_2)$ around 
$\omega_p/2$. This yields
\beq 
\Delta k(\omega_1,\omega_p -\omega_1) \approx  (\omega_1 - \omega_p/2)\times J,
\eeq
where $J := k^{\prime}_{{\text{idler}}}(\omega_p/2) - k^{\prime}_{{\text{signal}}}(\omega_p/2)$. The idler pass band function will be approximated as $\eta(\omega) =  e^{-(\omega - \omega_p /2)^2/\sigma^2}$, where $\sigma > 0$ is a real parameter. We will also approximate $\text{sinc}(x) \approx e^{-\gamma x^2}$, with $\gamma = 0.193 \hdots$ With these approximations, the $j$th input state is given by
\beq
\label{eq:jin}
\ket{\Psi}_{j} \approx \tilde{{\cal{C}}} \int d \omega e^{-\beta(\omega -\omega_p /2)^2} e^{2 \pi i \omega(t_j +t^{\prime})}
\hat{s}^{\dagger}(\omega) \ket{\text{vac}},
\eeq
with the parameters $\beta := 1/\sigma^2 - \gamma {\cal{L}}^2 J^2/4$, $t^{\prime} := -{\cal{L}} J/4 \pi$, and $\tilde{{\cal{C}}} :=  {\cal{C}}^* e^{-2 \pi i \omega_p t_j} e^{i {\cal{L}} J \omega_p/4}$. The time-dependent mode operators are related to their frequency-dependent counterparts via the standard Fourier transform $\hat{b}(t) = \int d\omega \hat{b}(\omega)e^{-2\pi i \omega t}$. This relation holds since we are assuming narrow band fields \cite{Blow}. The Fourier transform of Eq.~(\ref{eq:jin}) has the form
\bea
\ket{\Psi}_{j} &=&  {\cal{M}} \sqrt{\frac{\pi}{\beta}} \int dt e^{-2 \pi i t (\omega_p/2)} e^{-\pi^2 t^2 /\beta} \nonumber \\
&& \times \hat{s}^{\dagger}
(t_j +t^{\prime} - t) \ket{\text{vac}}, 
\eea
where ${\cal{M}}$ is a new normalization constant. Upon detection of the $j$th idler at time $t=t_j$, a photon is injected via mode 
$\hat{s}$ with a temporal distribution that oscillates with frequency $\omega_p/2$ under the 
Gaussian envelope centered at $t_j+ t^{\prime}$. The additional time advance/delay $t^{\prime}= -{\cal{L}} J/4 \pi$ depends on the characteristics of the nonlinear medium supporting the down-conversion process. 
We will assume that $t^{\prime}$ has been accounted for by using compensators outside of the crystal and 
henceforth neglect its contribution.

The time-dependent mode operators are related by 
\bea
\label{modes}
\hat{g}(t)&=&\frac{1}{\sqrt{2}}(\hat{a}(t)+i\hat{s}(t)), \;\;
\hat{u}_1(t)=\frac{1}{\sqrt{2}}(\hat{s}(t)+i\hat{a}(t)), \nonumber \\
\hat{u}_2(t)&=&e^{i\phi_1}\hat{u}_1(t-t_L), \;\;
\hat{u}_3(t)=e^{i\phi_2}\hat{u}_2(t-t_T), \nonumber \\
\hat{w}_1(t)&=&\frac{1}{\sqrt{2}}(e^{i(\phi_1+\phi_2)}\hat{g}(t-t_B)+i\hat{u}_3(t-t_R)), \nonumber \\
\hat{w}_2(t)&=&\frac{1}{\sqrt{2}}(e^{i(\phi_1+\phi_2+\pi/2)}\hat{g}(t-t_B)+\hat{u}_3(t-t_R)). 
\eea
In these expressions $t_L,t_T,t_R$, and $t_B$ denote the time it takes for a photon to travel through arms $L,T,R$, and $B$ respectively. Under normal operating conditions $t_L=l_L/c, t_T=l_T/c,t_R=l_R/c$, and 
$t_B=(l_L+l_T+l_R)/c$. We have also included the phase shifts which will be necessary for the prevention of certain intrusion attacks.
For the present discussion, we may set 
$\phi_1=\phi_2=0$.

Using the relations above, one can calculate the output flux at $\hat{w}_1(t)$ to be $\langle \hat{w}_1^{\dagger}(t) \hat{w}_1(t) \rangle_{normal} = \langle \hat{s}^{\dagger}(t-t_B) \hat{s}(t-t_B) \rangle $. This equation can be readily evaluated for $\ket{\Psi}_{j}$. 
We obtain
\beq
\label{eq:flux2}
\langle \hat{w}_1^{\dagger}(t) \hat{w}_1(t) \rangle_{normal} =  \frac{\pi |{\cal{M}}|^2}{\beta} {\text{exp}}{[-2 \pi^2 (t_j+t_B-t)^2/\beta]}.
\eeq
As expected, under normal conditions $\langle \hat{w}_1^{\dagger}(t) \hat{w}_1(t) \rangle$ 
is peaked at time $t= t_j+t_B$. Requiring $\int dt \langle \hat{w}_1^{\dagger}(t) \hat{w}_1(t) \rangle_{normal} =1$ 
yields ${\cal{M}} =(2\beta/\pi)^{1/4}$. With the state $\ket{\Psi}_{j}$ 
properly normalized, we may calculate the expectation value of the photon number arriving at $\hat{w}_1$ within 
the resolving time $T_R$ of the detector. Conditioned on the $j$th herald, the detector will make a measurement centered at time $t=t_j+t_B$.  We expect to find
\bea
{\cal{I}}_0 &:=& \int_{t_j+t_B-T_R/2}^{t_j+t_B+T_R/2} d\tau \langle \hat{w}_1^{\dagger}(\tau) \hat{w}_1(\tau) \rangle_{normal} \nonumber \\
&=& \text{Erf}\left(T_R  \pi/\sqrt{2 \beta} \right),
\eea
with the definition $\text{Erf}(x) := \frac{2}{\sqrt{\pi}}\int_0^x dz e^{-z^2}$. The error function $\text{Erf}\left(T_R  \pi/\sqrt{2 \beta} \right)$ increases from zero to one as the factor $T_R/\sqrt{\beta} >0$ increases. For $T_R \ge \sqrt{\beta}$, ${\cal{I}}_0  \approx 1$. Recall that $\beta = 1/\sigma^2 - \gamma {\cal{L}}^2 J^2/4$$\sigma$ and $\sigma$ determines the pass-band width of the idler detector. As $\sigma$ decreases, the idler frequency pass-band becomes tighter resulting in a broader temporal spread of the signal photon. Therefore, a smaller value for 
$\sigma$ requires a larger value of the resolving time $T_R$ to capture the entire signal amplitude.

\begin{figure}[!h]
\includegraphics[width=.4\textwidth]{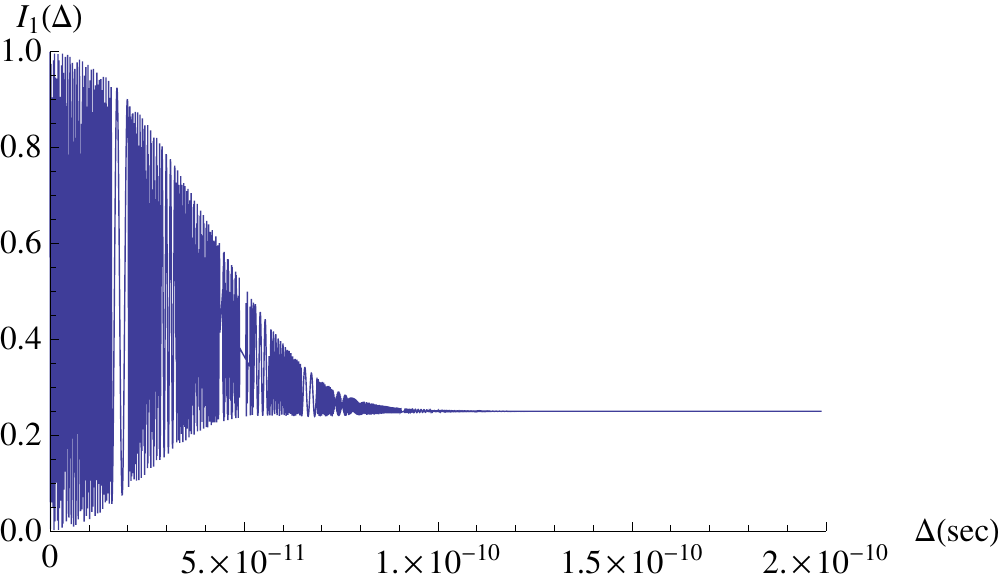}
\caption{Plot of ${\cal{I}}_1$ as a function of $\Delta$. We have taken the pump wavelength to be $\lambda=400$ nm and have 
set $T_R=\sqrt{\beta}=.1$ ns. The intensity drops to 1/4 of the normal value around $\Delta = .1$ ns, corresponding to a 
path length increase of $\approx$ 1 cm. The expectation values $\langle \hat{w}_1^{\dagger} \hat{w}_1^{\:} \rangle_{SI}$ and 
 $\langle \hat{w}_2^{\dagger} \hat{w}_2^{\:} \rangle_{SI}$ oscillate sinusoidally for small values of $\Delta$. Once the diversion length becomes larger than the coherence length, the detection interval captures amplitudes propagating along the bottom arm alone. No interference occurs beyond this point and $\langle \hat{w}_1^{\dagger} \hat{w}_1^{\:} \rangle_{SI}=\langle \hat{w}_2^{\dagger} \hat{w}_2^{\:} \rangle_{SI}=1/4$ for 
increasing $\Delta$.}
\label{fig:NI}
\end{figure}

\emph{Intrusion detection.---}
A clever intruder 
may attempt to fool the security device using path diversion or intercept-resend techniques. Let us first examine the case where a path redirection occurs along the same arm, i.e., the two points of contact to 
the original beam lie on either $L,T,$ or $R$. Due to the straight line geometry of each branch, a diversion such as this will  necessarily increase the optical path length. Using Eq.~(\ref{modes}), we may calculate  $\langle \hat{w}^{\dagger}_1(t)\hat{w}_1(t)\rangle_{SI}$ after the path length has been increased due to a side intrusion (SI): 
\bea
\langle \hat{w}_1^{\dagger}(t) \hat{w}_1(t) \rangle_{SI} &=& \frac{1}{4} \langle \left[ \hat{s}^{\dagger}(t-t_B) +e^{-i \xi}
\hat{s}^{\dagger}(t-(t_B+\Delta)) \right] \nonumber \\
&&\!\!\!\times \left[ \hat{s}(t-t_B) +e^{i\xi}
\hat{s}(t-(t_B+\Delta)) \right] \rangle
\eea
where $\Delta$ represents the additional flight time introduced as a result of the redirection, i.e., $\Delta=t_L+t_T+t_R-t_B$. The 
additional phase factor $\xi = 2\pi  c \Delta/\lambda$ results from the additional path length, where $c$ is the speed of light and 
$\lambda$ is the signal photon wavelength. For intrusions of this sort, the measured intensity is calculated to be
\begin{widetext}
\bea
\label{eq:one}
{\cal{I}}_1 := \int_{t_j+t_B-T_R/2}^{t_j+t_B+T_R/2} d\tau \langle \hat{w}_1^{\dagger}(\tau) \hat{w}_1(\tau) \rangle_{SI} 
&=& \frac{1}{8} 
\left[2 \:{\cal{I}}_0  +\text{Erf}\left[\frac{\pi(2 \Delta +T_R)}{\sqrt{2\beta}}  \right] -
\text{Erf}\left[\frac{\pi(2 \Delta -T_R)}{\sqrt{2\beta}}  \right] \right. \nonumber \\
&& \!\!\!\!\!\!\!\!  \left. + 2e^{-\pi^2 \Delta^2/2\beta} \cos{(\pi \Delta \omega_p-\xi)} \left(  \text{Erf}\left[ \frac{\pi(\Delta+T_R)}{\sqrt{2 \beta}}
\right] -\text{Erf}\left[ \frac{\pi(\Delta-T_R)}{\sqrt{2 \beta}}      \right]      \right) \right] 
\eea
\end{widetext}
In Fig.~\ref{fig:NI}, we plot an example of the intensity drop as a function of the intrusion parameter $\Delta$. In this example, we assume a pump beam centered at 400 nm and a detector resolving time of $T_R=\sqrt{\beta}=.1$ ns. We find 
that an additional path length increase $\geq 1$ cm results in a value of ${\cal{I}}_1$ equal to ${\cal{I}}_0/4$. 

For completeness, we must consider the unlikely situation where a photon amplitude propagating through the long path 
matches up with a short path amplitude associated with a different photon emitted from a later down-conversion. The two-photon 
input state corresponding to the $j$th and $k$th heralds is 
\bea
\ket{\Phi}_{j,k} &=&  \sqrt{\frac{2\pi}{\beta}} \int d \chi_1 d \chi_2 e^{-\pi i (\chi_1 + \chi_2) \omega_p} 
e^{-\pi^2 (\chi_1^2 +\chi_2^2) /\beta} \nonumber \\ 
&& \times \hat{s}^{\dagger}
(t_j - \chi_1)\hat{s}^{\dagger}(t_k - \chi_2) \ket{\text{vac}}. 
\eea
In what follows we assume the condition $t_j-t_k \geq \sqrt{\beta}$ is satisfied. We can enforce this condition with 
high probability by adjusting the pump strength. We then calculate  
\beq
\label{eq:Two}
\langle \hat{w}_1^{\dagger}(t) \hat{w}_1(t) \rangle_{j,k} = || \hat{w}_1(t)\ket{\Psi}_{j}||^2 +|| \hat{w}_1(t)\ket{\Psi}_{k}||^2.
\eeq
The expectation value of $\hat{w}_1^{\dagger}(t) \hat{w}_1(t)$ for the two-photon state $\ket{\Phi}_{j,k}$ is the sum 
of the expectation values for $\ket{\Psi}_{j}$ and $\ket{\Psi}_{k}$. The advantage of using a random photon source comes from the unlikeliness of witnessing a continual sequence of emissions with temporal separation $T= c\Delta m$, $m \in \mathbb{N}$. Therefore, in a diversion attack of this sort we expect the output flux to drop to 1/4 of the normal operating value for path diversion lengths necessary for human scale entry. When $\Delta = 0$, the integral of Eq.~(\ref{eq:Two}) over all time yields
$\int dt \langle \hat{w}_1^{\dagger}(t) \hat{w}_1(t) \rangle_{j,k} =2$. As expected, both photons emerge 
from $\hat{w}_1$ when there is no intrusion.

Now consider the case where a diversion occurs between separate branches, e.g., from $L$ to $T$. Since the optical path length inside the 
secured region could be determined using a probe signal, it would be possible 
for an intruder to perfectly match the original fence length in this case. Furthermore, since the sum of the length of the two legs 
of a right triangle is greater than the length of its hypotenuse, it would be possible for an intruder to match the original fence length while also providing sufficient entry room for an intrusion to take place. For this reason, we include random phase 
shifts $\phi_i$ at the corners of the perimeter. The sequence of phase shift values in 
a particular location can be generated before activation and stored as a string 
$\phi_i = (\phi_{i,1},\phi_{i,2},\phi_{i,3},\hdots,\phi_{i,N} )$ $(i=1,2)$. Suppose the length of the diverted 
path perfectly matches the original path length. In general, each diverted signal photon experiences an additional phase shift of $e^{i \phi_{int}}$. The mode relation between arms $L$ and $T$ is now given by $\hat{u}_2(t)=e^{i\phi_{int}}\hat{u}_1(t-t_L)$. For a cross intrusion (CI), we find the following expectation value 
\bea
\langle \hat{w}_1^{\dagger}(t) \hat{w}_1(t) \rangle_{CI} &=& \sqrt{\pi/2\beta} \exp{[-2 \pi^2 (t_j+t_B-t)^2/\beta]} 
\nonumber \\ 
&&\times \left[1 + \cos{(\phi_{int}-\phi_1)}\right], \
\eea
which reduces to Eq.~(\ref{eq:flux2}) when $\phi_{int} = \phi_1$. For a fixed $\phi_1$, the average 
output flux drops to half of the normal rate
\beq
\frac{1}{2\pi} \int_{\phi_{int}=0}^{2\pi} \int_{t=-\infty}^{\infty}dt 
d\phi_{int} \langle \hat{w}_1^{\dagger}(t) \hat{w}_1(t) \rangle_{CI}  = \frac{1}{2}.
\eeq \newline
Alternatively, we could generate phase shift values at the corners using quantum random number generators (QRNGs). Suppose we remove the phase shifts $\phi_1+\phi_2$ from the bottom arm and restrict the phase shift values at the corners to the set $\{ 0,\pi \}$. As each QRNG randomly selects between the  values $\phi_i=0,\pi \;(i=1,2)$ the expectation values for $\hat{w}_1^{\dagger}(t) \hat{w}_1(t)$ and $\hat{w}_2^{\dagger}(t) \hat{w}_2(t)$ change accordingly. Under normal operation 
$\hat{s} \rightarrow \frac{1}{2}\left[e^{-i(\phi_1+\phi_2)}(\hat{w}_2-i\hat{w}_1)-(\hat{w}_2+i\hat{w}_1)    \right],$
where we have suppressed the time dependence for brevity. The four combinations for $(\phi_1,\phi_2 )$ lead to the 
detection events  $(0,0 ),(\pi,\pi) \Rightarrow \hat{w}_1$ and $(0,\pi ),(\pi,0 ) \Rightarrow \hat{w}_2$. We can achieve relativistic security \cite{Jeffrey} by broadcasting the values $(\phi_1,\phi_2 )$ after the detections at 
$\hat{w}_1$ and $\hat{w}_2$ have already been made. In this way we eliminate the previous requirement of "keeping secrets". In this protocol, the 
sequence of detections at $\hat{w}_1$ and $\hat{w}_2$ are recorded and later compared to the expected values after the 
broadcast has taken place.

We now consider the intercept-resend approach.  In this case, the intruder constructs a device that injects a single photon (vacuum) into the fence branch if a photon is (is not) detected. In either case, the process of measuring the state of the perimeter ultimately destroys the superposition which was once present. A single photon will arrive at the second beam splitter, either from mode $\hat{g}$ or $\hat{u}_3$, but it will only have a 50$\%$ chance of arriving at the correct detector. This leads to an average output flux at $\hat{w}_1$ equal to ${\cal{I}}_0/2$. Any intrusion attempt that determines whether a photon was present in the fence branch will result in an average flux at $\hat{w}_1$ less than or equal to ${\cal{I}}_0/2$. Notably, strategies based on teleportation fall into this category.

\emph{Security tolerance.---}
The analysis above was based on a perfect, lossless implementation. In practice, this will not be the case. There will be some probability $0 \leq p_i \leq 1$ for detecting a photon at port $\hat{w}_i$ conditioned on the $j$th herald
\bea
\int_{t_j+t_B-T_R/2}^{t_j+t_B+T_R/2} d\tau \langle \hat{w}_i^{\dagger}(\tau) \hat{w}_i(\tau) \rangle_{normal} =p_i.
\eea
The sensor can operate in a realistic environment as long as $p_2$ remains much smaller than $p_1$ throughout normal, uninterrupted operation. These probabilities can be determined experimentally by examining the statistics associated with $N\gg1$ heralds. Let $n_j$ denote the number of clicks registered at port $\hat{w}_j$ during the $N$ 
detection intervals. (We assume no intrusion at this stage.) We can then define $p_j := n_j/N$ to be our expectation values. We 
will also define the expected and measured averages over any $N$ future heralding events by 
\bea
\label{eq:exp}
{\cal{A}}_{exp,i} &:=& \frac{1}{N}\sum_{j=1}^N \int_{t_j+t_B-T_R/2}^{t_j+t_B+T_R/2} d\tau \langle \hat{w}_i^{\dagger}(\tau) \hat{w}_i(\tau) \rangle_{normal}, \nonumber \\
\label{eq:meas}
{\cal{A}}_{mea,i} &:=& \frac{1}{N}\sum_{j=1}^N\int_{t_j+t_B-T_R/2}^{t_j+t_B+T_R/2} d\tau \langle \hat{w}_i^{\dagger}(\tau) \hat{w}_i(\tau) \rangle_{mea}, 
\eea
with ${\cal{A}}_{exp,i} =p_i$. For some values $\Gamma_i \in \mathbb{R}$, we have ${\cal{A}}_{exp,i} - {\cal{A}}_{mea,i} := \Gamma_i \;{\cal{I}}_0.$
In an ideal setting, with no photon loss or dephasing, and no intrusion, $\Gamma_i=0$ for both $i=1,2$. 
In a realistic setting, typical values for $\Gamma_i$ associated with specific intrusion techniques would need to be determined experimentally. Once they were identified, a user could specify security tolerances $\epsilon_i \geq 0$ that prompt an alarm when 
$|\Gamma_i| \geq \epsilon_i$. Alternatively, the user could define $0<\nu \leq 1$ to 
be the security tolerance and then constantly measure and update the quantity $\Theta := \frac{n_1}{n_1+n_2}$ for 
a predetermined value of $N$. When $\Theta < \nu$, an alarm is set signaling an intrusion.

\emph{Conclusion.---}
We have presented a method for detecting intrusion across an optically defined perimeter using the MZI effect. 
In an ideal setting, any human-scale intrusion attempt will lead to a decreased value of the output flux 
emitted from the "bright" port of a balanced Mach-Zehnder interferometer by at least half of the normal operating value. In a 
realistic setting, the interferometer can still function as an intrusion detection device as long as a large 
discrepancy remains between the normal flux values of the two outputs. This work complements recent studies related to the emerging 
field of quantum-based security \cite{Humble,Malik,Anisimov} and closely resembles a similar approach to realizing intrusion detection using quantum interference \cite{Anisimov}. However, we assume here that the location of the tripwire is publicly known, e.g., interlaced with an ordinary fence via optical fibers. We expect additional work will provide greater clarity into the sensing power of this approach, especially with respect to actively tampered optical seals.

\emph{Acknowledgements.---}
This work was sponsored by the Defense Threat Reduction Agency. This work was performed at Oak Ridge National Laboratory, operated by UT-Battelle for the U.S. Department of energy under Contract No. DE-AC05- 00OR22725. The work has been authored by a contractor of the U.S. Government. Accordingly, the U.S. Government retains a nonexclusive, royalty-free license to publish or reproduce the published form of this work, or to allow others to do so for U.S. Government purposes.



\end{document}